\documentclass[twocolumn]{aastex631}

\newcommand\VVSco{V*~V1042~Sco} 

\newcommand{\Fopt}{F_{\nu,{\rm opt}}}
\newcommand{\Fradio}{F_{\nu,{\rm radio}}}
\shorttitle{An optical counterpart search for FRB~20190520B}
\shortauthors{Niino et al.}

\begin{document}

\title{
Deep simultaneous limits on optical emission from FRB~20190520B 
by 24.4~fps observations with Tomo-e Gozen 
}

%% you desire. In v6.31 you can include a footnote in the title.

\correspondingauthor{Yuu Niino} 
\email{yuuniino@ioa.s.u-tokyo.ac.jp} 

\author[0000-0002-0786-7307]{Yuu Niino} 
\affil{Institute of Astronomy, Graduate School of Science, 
        The University of Tokyo, 2-21-1, Osawa, Mitaka, Tokyo 181-0015, Japan}
\affil{Research Center for the Early Universe, Graduate School of Science, 
        The University of Tokyo, Bunkyo-ku, Tokyo 113-0033, Japan}

\author{Mamoru Doi} 
\affil{Institute of Astronomy, Graduate School of Science, 
        The University of Tokyo, 2-21-1, Osawa, Mitaka, Tokyo 181-0015, Japan}
\affil{Research Center for the Early Universe, Graduate School of Science, 
        The University of Tokyo, Bunkyo-ku, Tokyo 113-0033, Japan}
\affil{Kavli Institute for the Physics and Mathematics of the Universe (WPI), 
        The University of Tokyo Institutes for Advanced Study, 
        The University of Tokyo, 5-1-5 Kashiwanoha, Kashiwa, Chiba 277-8583, Japan}

\author[0000-0002-8792-2205]{Shigeyuki Sako} 
\affil{Institute of Astronomy, Graduate School of Science, 
        The University of Tokyo, 2-21-1, Osawa, Mitaka, Tokyo 181-0015, Japan}

\author[0000-0001-5797-6010]{Ryou Ohsawa} 
\affil{Institute of Astronomy, Graduate School of Science, 
        The University of Tokyo, 2-21-1, Osawa, Mitaka, Tokyo 181-0015, Japan}

\author[0000-0002-2721-7109]{Noriaki Arima} 
\affil{Institute of Astronomy, Graduate School of Science, 
        The University of Tokyo, 2-21-1, Osawa, Mitaka, Tokyo 181-0015, Japan}
\affil{Department of Astronomy, Graduate School of Science, 
        The University of Tokyo, 7-3-1, Hongo, Bunkyo-ku, Tokyo 113-0033, Japan}

\author[0000-0002-9092-0593]{Ji-an Jiang} 
\affil{Kavli Institute for the Physics and Mathematics of the Universe (WPI), 
        The University of Tokyo Institutes for Advanced Study, 
        The University of Tokyo, 5-1-5 Kashiwanoha, Kashiwa, Chiba 277-8583, Japan}
\affil{National Astronomical Observatory of Japan, National Institutes of Natural Sciences,  
        2-21-1, Osawa, Mitaka, Tokyo 181-0015, Japan} 

\author[0000-0001-8537-3153]{Nozomu Tominaga} 
\affil{National Astronomical Observatory of Japan, National Institutes of Natural Sciences,  
        2-21-1, Osawa, Mitaka, Tokyo 181-0015, Japan} 
\affil{Department of Astronomical Science, School of Physical Sciences, 
        The Graduate University of Advanced Studies (SOKENDAI), 
        2-21-1 Osawa, Mitaka, Tokyo 181-8588, Japan}
\affil{Department of Physics, Faculty of Science and Engineering, 
        Konan University, 8-9-1 Okamoto, Kobe, Hyogo 658-8501, Japan}
\affil{Kavli Institute for the Physics and Mathematics of the Universe (WPI), 
        The University of Tokyo Institutes for Advanced Study, 
        The University of Tokyo, 5-1-5 Kashiwanoha, Kashiwa, Chiba 277-8583, Japan}

\author[0000-0001-8253-6850]{Masaomi Tanaka} 
\affil{Astronomical Institute, Tohoku University, Sendai 980-8578, Japan} 
\affil{Division for the Establishment of Frontier Sciences, 
        Organization for Advanced Studies, Tohoku University, Sendai 980-8577, Japan} 

\author[0000-0003-3010-7661]{Di Li} 
\affil{National Astronomical Observatories, Chinese Academy of Sciences, Beijing 100101, China} 
\affil{NAOC-UKZN Computational Astrophysics Centre, 
    University of KwaZulu-Natal, Durban 4000, South Africa} 

\author[0000-0001-6651-7799]{Chen-Hui Niu} 
\affil{National Astronomical Observatories, Chinese Academy of Sciences, Beijing 100101, China} 

\author[0000-0002-9390-9672]{Chao-Wei Tsai} 
\affil{National Astronomical Observatories, Chinese Academy of Sciences, Beijing 100101, China} 

\author[0000-0003-4578-2619]{Naoto Kobayashi} 
\affil{Kiso Observatory, Institute of Astronomy, School of Science, The University of Tokyo,
        10762-30 Mitake, Kiso-machi, Kiso-gun, Nagano 397-0101, Japan} 

\author{Hidenori Takahashi} 
\affil{Kiso Observatory, Institute of Astronomy, School of Science, The University of Tokyo,
        10762-30 Mitake, Kiso-machi, Kiso-gun, Nagano 397-0101, Japan} 

\author{Sohei Kondo} 
\affil{Kiso Observatory, Institute of Astronomy, School of Science, The University of Tokyo,
        10762-30 Mitake, Kiso-machi, Kiso-gun, Nagano 397-0101, Japan} 

\author{Yuki Mori} 
\affil{Kiso Observatory, Institute of Astronomy, School of Science, The University of Tokyo,
        10762-30 Mitake, Kiso-machi, Kiso-gun, Nagano 397-0101, Japan} 

\author{Tsutomu Aoki} 
\affil{Kiso Observatory, Institute of Astronomy, School of Science, The University of Tokyo,
        10762-30 Mitake, Kiso-machi, Kiso-gun, Nagano 397-0101, Japan} 

\author[0000-0003-1260-9502]{Ko Arimatsu} 
\affil{Hakubi Center / Astronomical Observatory, Graduate School of Science, Kyoto University
        Kitashirakawa-Oiwakecho, Sakyo-ku, Kyoto 606-8502, Japan} 

\author[0000-0001-5903-7391]{Toshihiro Kasuga} 
\affil{National Astronomical Observatory of Japan, National Institutes of Natural Sciences,  
        2-21-1, Osawa, Mitaka, Tokyo 181-0015, Japan} 

\author[0000-0002-1873-3494]{Shin-ichiro Okumura} 
\affil{Japan Spaceguard Association, Bisei Spaceguard Center, 1716-3 Okura, 
        Bisei-cho, Ibara, Okayama 714-1411, Japan} 


\begin{abstract}
We conduct 24.4~fps optical observations 
of repeating Fast Radio Burst (FRB) 20190520B using Tomo-e Gozen, 
a high-speed CMOS camera mounted on the Kiso 105-cm Schmidt telescope, 
simultaneously with radio observations carried out using 
the Five-hundred-meter Aperture Spherical radio Telescope (FAST). 
We succeeded in the simultaneous optical observations 
of 11 radio bursts that FAST detected.
However, no corresponding optical emission was found. 
The optical fluence limits as deep as 0.068 Jy ms 
are obtained for the individual bursts (0.029 Jy ms on the stacked data)  
corrected for the dust extinction in the Milky Way. 
The fluence limit is deeper than those obtained 
in the previous simultaneous observations 
for an optical emission with a duration $\gtrsim 0.1$ ms. 
Although the current limits on radio--optical 
spectral energy distribution (SED) of FRBs are not constraining, we show that SED models 
based  on observed SEDs of radio variable objects such as optically detected pulsars, 
and a part of parameter spaces of theoretical models in which FRB optical emission 
is produced by inverse-Compton scattering in a pulsar magnetosphere 
or a strike of a magnetar blastwave into a hot wind bubble, 
can be ruled out once a similar fluence limit as in our observation 
is obtained for a bright FRB with a radio fluence $\gtrsim 5$ Jy ms. 
\end{abstract} 


% \keywords{radio continuum: general --- methods: observational}
\keywords{Radio transient sources (2008) --- Optical observation (1169) --- Time domain astronomy (2109)}


\section{Introduction} 

A Fast Radio Burst (FRB) is a transient astronomical 
object observed at $\sim$ 1 GHz frequency with a typical duration 
of several milliseconds, whose origin is not yet known 
\citep[e.g.,][]{Lorimer:2007a, Thornton:2013a}. 
Roughly 600 FRB sources have been discovered so far, 
among which more than 20 FRB sources are known 
to produce bursts repeatedly (repeating FRBs), 
while other FRB sources do not show any repetition (non-repeating FRBs). 
FRBs have large dispersion measures (hereafter DMs) 
that exceed the expected amounts within the Milky Way (MW) in their direction. 
Their large DMs suggest that FRBs are extragalactic objects. 
Although various theoretical models have been proposed 
\citep[e.g.,][see \citeauthor{Platts:2019a}~\citeyear{Platts:2019a} 
for a review]{Totani:2013a, Kashiyama:2013a, 
Popov:2013a, Falcke:2014a, Cordes:2016b, Zhang:2017a}, 
observational evidence that confirms or rejects those models is still lacking. 

Majority of the currently known FRBs have been discovered by widefield 
radio telescopes with typical localization accuracy of $\gtrsim$ 10 arcmin, 
and hence it is challenging to identify their counterparts or host galaxies in most of the cases. 
Currently, identifications of FRB host galaxies, 
and hence distance measurements that are independent of DM, 
have been achieved only for $\sim$ 20 FRBs among the $\sim$ 600 FRBs. 
Distances of other FRBs are estimated from their DMs assuming 
that the DMs in excess of the expected MW component 
arise mostly from the inter-galactic medium (IGM), 
and considered to be widely distributed over a redshift range $z \sim 0.1$--2. 
In cases where a host galaxy of an FRB is known, 
the distance estimate from their DMs are mostly consistent 
with the redshifts of the host galaxies \citep{Macquart:2020a}, 
however, there is a case of FRB~20190520B (also referred as FRB~190520B) 
in which the DM indicates much larger distance than inferred from the redshift \citep{Niu:2021a}. 

Discovery of a counterpart of a mysterious astronomical object 
in other observational passband often revolutionize 
our understanding on the nature of the object. 
No clear transient counterpart of an FRB has been found in any wavelength  
despite the counterpart searches carried out in various passbands 
and timescales \citep[e.g.,][]{Petroff:2015a, Niino:2018a, Tominaga:2018a}, 
except the case of FRB~200428A, an FRB like burst from a galactic magnetar SGR~1935+2154 
\citep{The-CHIME/FRB-Collaboration:2020a, Bochenek:2020a},  
which was detected simultaneously with an extraordinary X-ray flare 
from the same magnetar \citep{Ridnaia:2021a}. 
Recently, \citet{Li:2022a} reported that an optical transient event 
AT2020hur spatially coincided with repeating FRB 180916B within 1 arcsec, 
although it is not clear weather AT2020hur is an emission 
from the same object as the repeating FRB source. 

A few pulsars are detected both in radio and optical \citep[e.g.,][]{Danilenko:2011a}. 
It is possible that FRBs have optical emission if they are produced 
by a similar emission mechanism as that of pulsars, 
which, in itself, is still poorly understood. 
\citet[][hereafter Y19]{Yang:2019a} has shown that 
an detectable optical emission that accompanies an FRB 
can be produced by inverse-Compton scattering (IC). 
It is also possible that a blastwave from a magnetar cause 
a bright optical flare when it collides into a hot wind bubble produced 
by a previous flare \citep[][hereafter B20]{Beloborodov:2020a}. 
The existence of complex magneto-ionic environments near a few repeating FRBs, 
which possibly supports the blastwave scenario, has been indicated 
by the analysis of observed frequency-dependent polarization \citep{Feng:2022a}. 
However, search for an optical FRB counterpart on a timescale 
that is comparable to the timescale of an FRB 
is especially challenging because most of the existing observing facilities 
require longer observing timescale than a few seconds 
\citep[see e.g.,][for searches of an optical emission from an FRB 
on longer timescales]{Andreoni:2020a, Xin:2021a, Kilpatrick:2021a}. 

There is a few optical upper limits with a sub-second timescale 
that were obtained simultaneously with several radio bursts from FRB~121102, 
i.e. the first repeating FRB source discovered, 
using telescopes equipped with an electron-multiplying CCD camera which enables 
observations in a sub-second time resolution \citep{Hardy:2017a, Karpov:2019a}  
or using gamma-ray Cherenkov telescopes as optical facilities \citep{MAGIC-Collaboration:2018a}. 
However no optical counterpart has been detected so far. 
Furthermore, a search of optical emission 
from other repeating FRBs with a sub-second timescale is missing 
despite the growing number of repeating FRB sources discovered. 

In this paper, we present searches for optical emission from repeating FRB~20190520B 
with a 24.4~fps observation using a high-speed CMOS camera, 
Tomo-e Gozen \citep{Sako:2018a}, mounted on the Kiso 105-cm Schmidt telescope, 
which are conducted simultaneously with radio observations 
by the Five-hundred-meter Aperture Spherical radio Telescope (FAST). 
In section~\ref{sec:obs}, we describe our observation and data analysis. 
In section~\ref{sec:burstdata}, we investigate optical data that corresponds 
to the arrival time of the radio bursts detected during the observations, 
and discuss upper limits on the optical fluences of the bursts obtained by our observation. 
We summarize our conclusion in section~\ref{sec:conclusion}. 
 
\section{Observations} 
\label{sec:obs}

Repeating FRB~20190520B was discovered with FAST \citep{Nan:2011a, Li:2013a}. 
The initial detection was obtained during a drift-scan on 2019 May 20, 
as part of the Commensal Radio Astronomy FAST Survey \citep[CRAFTS,][]{Li:2018a}. 
Subsequent sub-arcsecond localization was obtained in July, 2020 
with the Karl G. Jansky Very Large Array (VLA). 
FRB~20190520B is characterized by its large DM $\sim 1200$ cm$^{-3}$~pc 
compared to the redshift of the host galaxy $z = 0.24$, 
and the association with a compact persistent radio source \citep{Niu:2021a}. 
The large DM indicates the DM component that arise from 
gas in the host galaxy reaches $\sim 900$ cm$^{-3}$~pc, 
which is nearly an order of magnitude higher than that in other FRB host galaxies 
and overwhelms the IGM component which can be used as a distance indicator. 
To investigate optical emission component of the bursts from FRB~20190520B, 
we conduct optical monitoring observations of the FRB using Tomo-e Gozen, 
simultaneously with observations of the same FRB by FAST. 

\begin{figure*}
\plotone{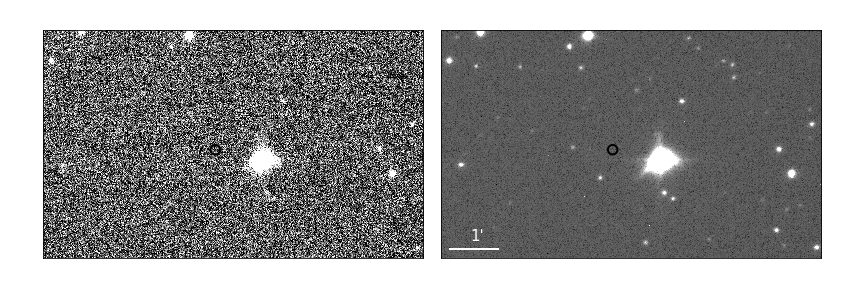} 
\caption{
The $7.9\times4.8$ arcmin$^2$ field image around 
the coordinate of FRB~20190520B (marked with a circle in each panel). 
The north is up and the east to the left. 
The left panel shows a single frame image with an integration time of 40.9 ms 
which corresponds to the DM corrected arrival time 
of the radio burst with the largest fluence during the observation 
\citep[burst ID P50 in][the first burst on August 14]{Niu:2021a}. 
The right panel shows the stacked image of the 1000 frames 
which are included in the FITS file that contains the frame shown in the left panel. 
\label{fig:field} 
}
\end{figure*}

\begin{figure*}
\plotone{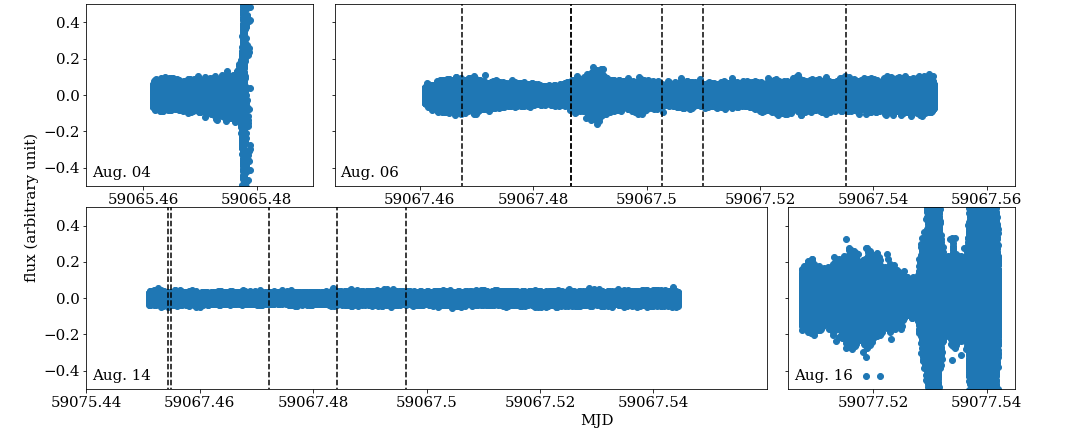} 
\caption{
The optical lightcurve obtained by the forced photometry 
of the coordinate of FRB~20190520B with an aperture radius of 6 arcsec. 
The DM corrected arrival times of the 11 radio bursts 
detected by FAST are shown with vertical dashed lines. 
\label{fig:lightcurve} 
}
\end{figure*}

We observed FRB~20190520B on the nights 
of 2020 August 4, 6, 14, and 16, using Tomo-e Gozen. 
Although the highest frame rate Tomo-e Gozen can achieve 
with full-frame readout ($2000\times1128$ pixels) is 2~fps, 
higher frame rate can be achieved with partial-frame readout.
For the observations of FRB~20190520B, 
we achieve a frame rate of 24.4~fps with 40.9 ms integration in each frame, 
with a partial readout of $400\times240$ pixels in each sensor 
($7.9\times4.8$ arcmin$^2$ with a pixel scale of 1.189 arcsec/pix) 
with the long side oriented in the East-West direction. 
Output data are written as 3D FITS files with each file containing a series of 1000 frames. 
The time gaps between the frames are as short as 0.1 ms, 
however there is a longer time gap of $\sim$ 1 sec 
in every 1000 frames (41 sec) during which the data is written to a FITS file. 

We do not use a filter for the observations, 
and the observing passband is determined 
by the spectral response function of the CMOS sensors, 
which covers a wavelength range $\sim$ 370--730 nm 
with a peak at 500 nm \citep{Kojima:2018a}. 
The timestamps of the Tomo-e Gozen data 
are GPS-synchronized, and have an accuracy of $< 1$ ms. 
Subtraction of bias and dark, and flat-fielding have been performed in a standard manner. 
The calibration data is obtained at the beginning of each night. 
A dark+bias image is generated by stacking 4500 frames (40.9 ms/frame) for each night, 
and each flat image is generated by stacking 180 dome-flat frames (1 sec/frame). 

A single frame image (40.9 ms exposure) and a stacked image of 1000 frames 
of the $7.9\times4.8$ arcmin$^2$ field around FRB~20190520B are shown in figure~\ref{fig:field}. 
No object is detected at the coordinate of FRB~20190520B. 
We perform forced photometry at the coordinate in each frame. 
The lightcurve obtained by the forced photometry is shown in figure~\ref{fig:lightcurve}. 
The arrival times of the radio bursts at the solar system barycenter are reported in \citet{Niu:2021a}. 
We convert the timestamps of the optical data to the time at the barycenter 
using the software package astropy \citep{Astropy-Collaboration:2013a}, 
in order to compare them with the arrival time of the radio bursts. 

Here we perform an approximated photometric calibration of all optical data, 
using a nearby bright star \VVSco\  located at $\sim$ 1 arcmin west 
of FRB~20190520B as a photometric standard. 
We assume that the flux of \VVSco\ does not vary 
during observation in each day, although \VVSco\ is known to be 
a pulsating K-type giant with a variability amplitude $\pm 0.1$ mag 
and a period of $\sim 30$ days \citep{Watson:2006a, Alfonso-Garzon:2012a}. 
We perform more accurate calibration of the data that correspond 
to the arrival times of the radio bursts in the next section. 
The observing condition was good on the night of August 14, while it was slightly cloudy on August 6.  
The conditions were unstable on the nights of August 4 and 16. 
Significant signal is not found above the noise level in the optical lightcurve. 

\begin{figure}
\centering
\includegraphics[width=8cm]{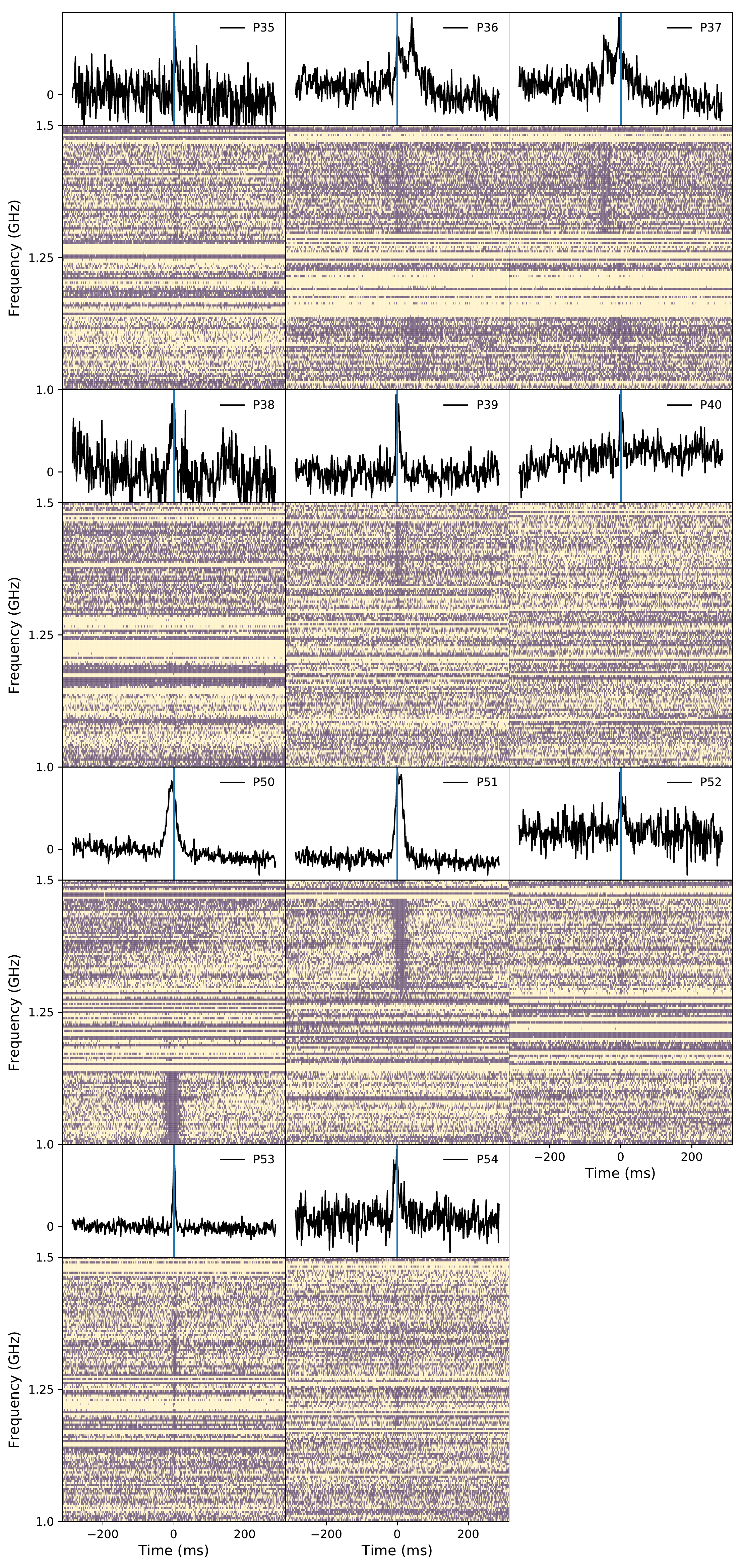}
% \plotone{Combine_pdf_v2.pdf}
\caption{
The dynamic spectra of the 11 radio bursts from FRB~20190520B 
detected by FAST during the simultaneous observations. 
The bursts with IDs P35--40 were detected on August 6, 
and those with IDs P50--54 were detected on August 14. 
The burst IDs are the same as those in \citet{Niu:2021a}. 
\label{fig:radioburst} 
}
\end{figure}

FAST detected 6 and 5 radio bursts during the simultaneous observations on August 6 and 14, respectively. 
In order to examine the optical image frames that corresponds to the arrival times of the radio bursts, 
we correct the DM effect on the arrival times that are reported at 1.5~GHz 
assuming the best estimate DM in each day as presented in \citet{Niu:2021a}. 
The DM corrected arrival times of the radio bursts 
and their dynamic spectra are shown 
in figure~\ref{fig:lightcurve} and figure~\ref{fig:radioburst}, respectively. 

\section{Optical data at the arrival times of the radio bursts} 
\label{sec:burstdata} 

\subsection{Photometric calibration} 

In order to perform photometric calibration of the optical data 
that correspond to the arrival times of the radio bursts 
independently of the variable star \VVSco, 
we use stars in the Pan-STARRS1 catalog \citep[][PS1]{Chambers:2016a} 
around the FRB~20190520B coordinate 
in the magnitude range $15.2 < r < 16.5$ as photometric standards. 
We perform photometry of the stars on stacked images 
of all frames contained in each FITS file (1000 frames), 
assuming that the photometric zeropoint does not vary 
during an exposure for 1000 frames, i.e. $\sim$ 41 sec. 
We use an aperture radius of 6 arcsec while the typical seeing size 
of the images is full width at half maximum (FWHM) $\sim$ 3--4 arcsec. 
The typical 5$\sigma$ limiting magnitudes of the stacked images 
are $\sim 18.0$ and 18.8 during the observations on August 6 and 14, respectively. 

To correct for the difference of passbands between Tomo-e Gozen and PS1, 
we compute broad band colors of the template stellar spectra provided by \citet{Kesseli:2017a}, 
by convoluting the template spectra with the spectral response functions of the CMOS sensors 
equipped in Tomo-e Gozen, and the $g-$, $r-$, $i-$band filters of PS1. 
We perform 2 dimensional second order polynomial fit 
of the broad band colors of the spectral templates 
with $g-m_{\rm T}$ color as a function of $g-r$ and $r-i$ colors, 
where $m_{\rm T}$ represents apparent magnitude in the Tomo-e Gozen passband. 
The result of the fitting is $g-m_{\rm T} = -0.1045+0.2938(g-r)\
+0.0852(r-i)-0.0076(g-r)(r-i)+0.2390(g-r)^2+0.2289(r-i)^2$, 
and the standard deviation of the fitting residuals is 0.025 mag. 

We compute the expected $m_{\rm T}$ of the stars 
around FRB~20190520B using the $g-$, $r-$, $i-$band magnitudes provided 
in the PS1 catalog and the color correlation derived above. 
We determine photometric zeropoints within each exposure 
comparing the photometric counts in the stacked images with the expected $m_{\rm T}$. 
The typical standard deviations of the zeropoint fit residuals 
are 0.10 mag and 0.05 mag on August 6 and 14, respectively, 
which we consider as the uncertainty of the photometric calibration. 

\subsection{Limits on optical fluences of the bursts} 

\begin{figure*}
\plotone{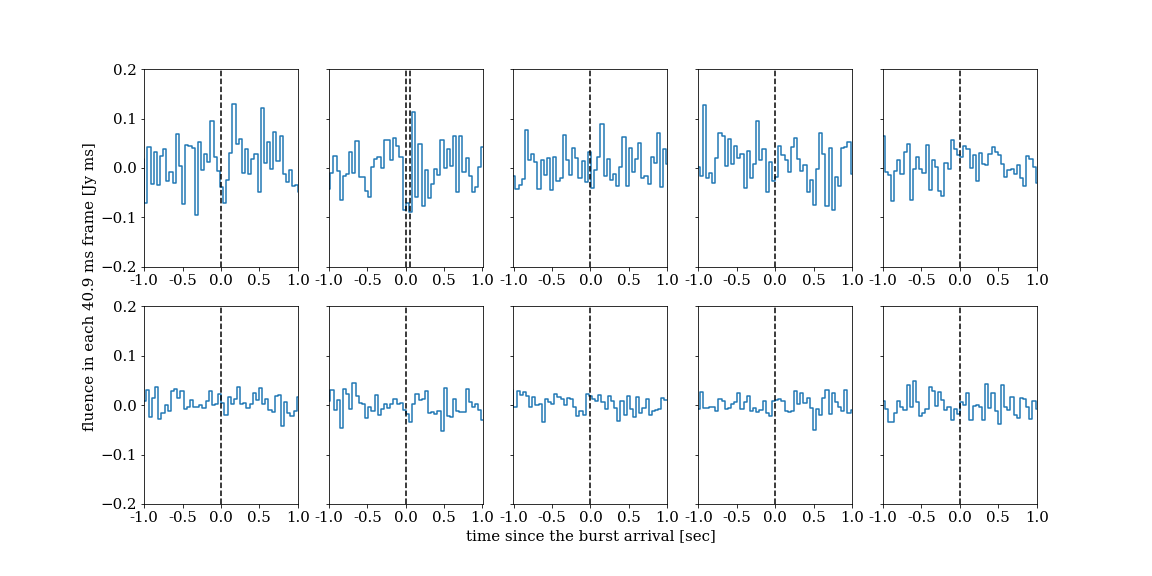} 
\caption{
The optical lightcurve within the time range $\pm 1.0$ sec of the burst arrival times. 
The second and third burst on August 6 are separated by only 50 ms in time, 
and hence shown in a same panel. 
\label{fig:closeup_lc} 
}
\end{figure*}

The photometrically calibrated optical lightcurve around 
the radio burst arrival times are shown in figure~\ref{fig:closeup_lc}. 
Any significant excess of optical flux is not found within $\pm 1.0$ sec of the burst arrival times. 
The limits on the optical fluences in the frames 
that correspond to the burst arrival times are estimated 
by performing point spread function (PSF) 
photometry at 1000 random positions in each frame, 
assuming Gaussian PSFs with FWHM $= 3.1$ arcsec on August 6, 
and 3.8 arcsec on August 14. 
We note that the data is read out from the CMOS sensor in a rolling shutter scheme, 
and hence the readout time is different in different positions in a single frame. 
We identify the frames corresponding to the burst arrival times 
using the readout time at the position of the FRB in each image. 

The 5$\sigma$ optical limits obtained with Tomo-e Gozen span 
$m_{\rm T} =$ 16.6--15.7 (or 0.86--1.88 mJy) 
which corresponds to the fluence range of 0.035--0.077 Jy ms 
with the integration time of 40.9 ms without correction for dust extinction. 
We stack the 11 frames that correspond to the burst arrival times 
in order to constrain the averaged optical flux of the 11 bursts, 
and obtain optical upper limit of $m_{\rm T} = 17.5$ (0.36 mJy) 
corresponding to the fluence of 0.015 Jy ms. 
We also stack $\pm 5$ frames and $\pm 25$ frames 
around the frames corresponding to the burst arrival times, 
i.e. 11 frames (0.45 sec) and 51 frames (2.1 sec) for each of the 11 bursts, 
in order to investigate optical fluence on longer timescales. 
By stacking the 121 frames and the 561 frames for the 11 bursts, 
we obtain fluence limits of 0.067 Jy ms and 0.176 Jy ms 
on the timescales of 0.45 sec and 2.1 sec, respectively. 
We allow duplicate frames in the stacking when multiple radio bursts 
are detected within the timescale considered. 

\begin{figure}
\centering
\includegraphics[width=9cm]{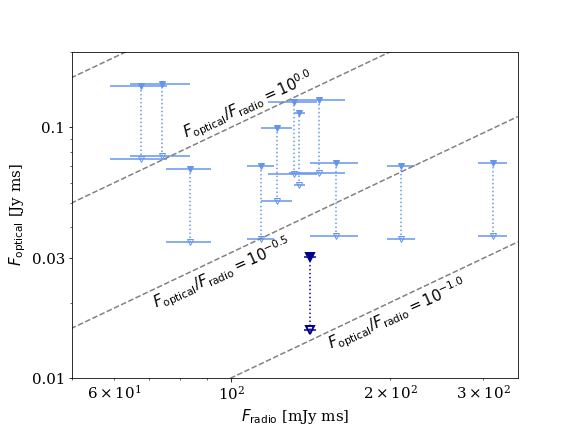}
% \plotone{fig_rad_opt_2.png} 
\caption{
Upper limits on the optical fluences of the bursts ($5\sigma$) as a function of the radio fluences.
The limits with and without the MW dust correction are filled and empty symbols, respectively. 
The thick datapoint in the bottom of the panel indicates 
the optical limit obtained by the stacking of the 11 frames that correspond 
to the arrival times of the radio bursts and the mean radio fluence of the 11 bursts. 
\label{fig:opt_limit} 
}
\end{figure}

The color excess in the MW in the direction of FRB~20190520B 
is $E_{B-V, {\rm MW}} = 0.25$ based on the color excess map 
by \citet[][https://www.ipac.caltech.edu/doi/ned/10.26132/NED5]{Schlafly:2011a}. 
We estimate the extinction in the Tomo-e Gozen passband ($A_{\rm T, MW}$) assuming flat $F_\lambda$ spectrum of a source 
as $10^{-0.4 A_{\rm T}} = \int\lambda\epsilon_{\rm T} 10^{-0.4 R_\lambda E_{B-V}}d\lambda/\int\lambda\epsilon_{\rm T} d\lambda$ 
where $\epsilon_{\rm T}$ is the spectral response function of the CMOS sensors
and $R_\lambda$ is the extinction law derived by \citet{Cardelli:1989a}, 
and obtain $A_{\rm T, MW} = 0.72$. 
Corrected for the extinction in the MW, the optical fluence limits 
are 0.068--0.149 Jy ms for individual bursts, 0.029 Jy ms for the stacked data, 
and 0.13 (0.34) Jy ms on the timescale of 0.45 (2.1) sec. 
The optical fluence limits are shown in figure~\ref{fig:opt_limit} as a function 
of radio fluence of each burst with and without the $A_{\rm T, MW}$ correction. 

It is also possible that optical emission from FRB~20190520B is affected 
by the dust extinction within the host galaxy of the FRB. 
Although the host galaxy dust extinction along the line-of-sight to the FRB is not known, 
the averaged extinction in the host galaxy can be estimated 
from the flux ratio of the emission lines in the Balmer series. 
The host galaxy of FRB~20190520B is a dwarf galaxy 
with high specific star formation rate \citep{Niu:2021a}. 
Spectroscopic observation of the host galaxy  
revealed $F_{\rm H_\alpha} = (23.9\pm0.3)\times10^{-17}$ erg cm$^{-2}$s$^{-1}$, 
and $F_{\rm H_\beta} = (6.2\pm0.3)\times10^{-17}$ erg cm$^{-2}$s$^{-1}$ 
corrected for the MW dust extinction (\citeauthor{Ocker:2022a}~\citeyear{Ocker:2022a}; Tsai et al., in prep.). 

Assuming the intrinsic line ratio of $F_{\rm H_\alpha}/F_{\rm H_\beta} = 2.86$ 
and the extinction law derived by \citet{Calzetti:2000a} 
which is often used for an actively star forming galaxy, 
the estimated color excess is $E_{B-V, {\rm host}} = 0.25$ in the rest frame of the host galaxy 
which corresponds to $A_{\rm T, host} = 0.94$ in the observer frame. 
However, we note that FRB~20190520B is located $\sim$ 1.3 arcsec 
offset ($\sim$ 5 kpc) from the center of the host galaxy, 
and the spatial structure of the emission lines in the host galaxy 
is not resolved in the spectroscopic observation. 
Hence the line-of-sight dust extinction to FRB~20190520B 
might be different from the extinction estimated from the Balmer lines. 

\subsection{Comparison to the previous simultaneous limits in optical} 

\citet{MAGIC-Collaboration:2018a} carried out a simultaneous multi-wavelength observation 
of repeating FRB~121102 using the Arecibo radio telescope and the MAGIC 
(Major Atmospheric Gamma Imaging Cherenkov) telescope, 
in which 5 radio bursts were detected. 
They put upper limits on the optical flux of the radio bursts 
using the Cherenkov telescope as an optical facility. 
By stacking the optical data around the burst arrival times, 
they achieved fluence limits of 0.0012, 0.0041, 0.012, and 0.017 Jy ms  
on timescales of 0.1, 1, 5, and 10 ms, respectively 
(5$\sigma$ in $U$-band without correction for the dust extinction). 
The MW color excess in the direction of FRB~121102 is $E_{B-V} = 0.68$, 
which corresponds to the extinction of $A_U = 3.3$ in $U$-band. 
When correction for the dust extinction is applied, 
the fluence limit obtained by MAGIC 
is 0.025, 0.086, 0.25, and 0.36 Jy ms depending on the timescale. 

\citet{Hardy:2017a} also carried out a simultaneous observation of FRB~121102, 
using the 100-m Effelsberg Radio Telescope and an electron-multiplying CCD camera, 
ULTRASPEC, mounted on the 2.4-m Thai National Telescope. 
They detected 13 radio bursts. Stacking the relevant optical image frames, 
they achieved the fluence limit of 0.046 Jy ms with a time resolution of $\sim 140$ ms  
at an observing wavelength of 767 nm without extinction correction. 
The MW color excess in the direction of FRB~121102 
($E_{B-V} = 0.68$) corresponds to $A_{767{\rm nm}} = 1.4$. 
Corrected for the MW dust extinction, the fluence limit is $0.17$ Jy ms. 

Thus, our fluence limit from the stacked data, 0.029 Jy ms with the timescale of 40.9 ms, 
is deeper than the previous optical fluence limits for an optical emission 
with a timescale $\gtrsim$ 0.1 ms which is typical of FRB durations in radio, 
while the observation by MAGIC puts a deeper limit for an emission with shorter duration. 
In the following sections, we discuss optical fluence limits 
corrected for dust extinction in the MW unless stated otherwise. 

% Thus the MW extinction corrected limit on the fluence ratio $\sim 10^{-3.3}$
% by our observation is $\sim 10^2$ times deeper than the previous one. 

% The radio bursts have fluences of $\sim$ 0.3--4 Jy ms and 
% corresponding to $\Fopt/\Fradio \lesssim 10^{-2.7}$ for a burst from FRB~121102. 

% Hence the limit of $\Fopt/\Fradio < 10^{-3.6}$ obtained for a burst 
% from FRB~20190520B \citep[burst ID P50 in][]{Niu:2021a} 
% by our observation is $\sim$ 8 times deeper than the previous limits. 

\subsection{Optical to radio fluence ratio} 

Here we discuss the limits on optical emission relative to the fluence of the radio bursts. 
The 11 radio bursts detected by FAST during the simultaneous observations 
have radio fluences ranging $\sim$ 70--310 mJy ms 
\citep[see][for the properties of each burst]{Niu:2021a}, 
and the optical limit to radio fluence ratio is $\Fopt/\Fradio < 10^{-0.64}$--$10^{0.33}$ 
for individual bursts (figure~\ref{fig:opt_limit}). 
Comparing the stacked optical limit ($< 0.029$ Jy ms) 
to the averaged radio fluence of the bursts (0.141 Jy ms), 
we obtain $\Fopt/\Fradio < 10^{-0.69}$. 

The averaged radio fluence of the 5 radio bursts detected 
by Arecibo in \citet{MAGIC-Collaboration:2018a} is $\sim$ 2 Jy ms, 
and hence their optical limit obtained from the stacked data, 
0.025--0.36 Jy ms depending on the timescale considered (0.1--10 ms), 
corresponds to $\Fopt/\Fradio \lesssim 10^{-1.9}$--$10^{-0.74}$. 
The median radio fluence of the radio bursts detected 
by the Effelsberg telescope in \citet{Hardy:2017a} is 0.6 Jy ms, 
with which their optical fluence limit of $< 0.17$ Jy ms corresponds to $\Fopt/\Fradio < 10^{-0.55}$. 
\citet{Chen:2020a} also discussed $\Fopt/\Fradio$ of FRBs 
in a statistical way based on wide field surveys 
in various observing wavelength ranging from optical to gamma-ray, 
assuming that the functional form of the fluence distribution 
of FRBs remains unchanged in any wavelength. 
They found that $\Fopt/\Fradio \lesssim 10^{-2.5}$ in optical, 
although this limit is an average over a wide survey area 
and it is difficult to correct for the dust extinction. 

\begin{figure}
\centering
\includegraphics[width=9cm]{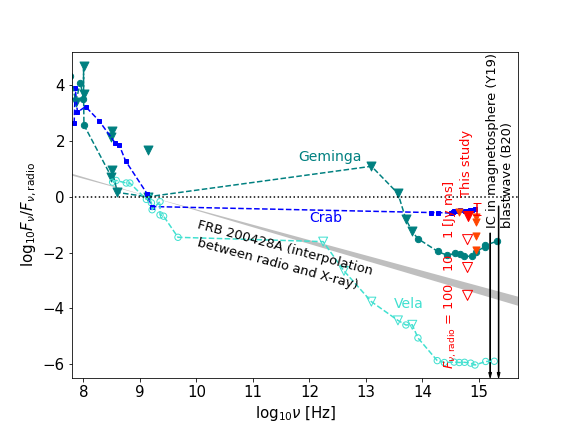}
\caption{
The limit on the optical to radio fluence ratio with correction for the extinction in the MW 
obtained by stacking the image frames that correspond to the burst arrival times 
(the filled downward triangle at log$_{10} \nu$ [Hz] = 14.8). 
The errorbar shown beside the limit represents the dust extinction 
in the host galaxy as estimated from the Balmer lines in the host galaxy spectrum, 
although it might be different from the dust extinction 
along the line-of-sight to FRB~20190520B. 
The empty downward triangles below the filled one indicate the limits 
that would be obtained by our optical fluence limit for a radio burst 
with radio fluence of 1, 10, and 100 Jy ms, from top to bottom. 
The smaller downward triangles shown on the left and right 
of the limits obtained in this study are the limits 
in \citet{Hardy:2017a} and \citet{MAGIC-Collaboration:2018a} 
with respect to the fluences of the radio bursts detected 
in their study ($\Fradio \sim$ 0.6 and 2 Jy ms, respectively). 
The limits by \citet{MAGIC-Collaboration:2018a} are shown 
for the timescales of 0.1, 1, 5, and 10 ms, from bottom to top. 
The filled squares represent the observed radio--optical SED of the Crab pulsar \citep{Buhler:2014a}. 
The SEDs of the Geminga and the Vela pulsar \citep{Danilenko:2011a} 
are shown with filled and empty circles (downward triangles for upper limits), respectively. 
The datapoints representing the same pulsars are connected with dashed lines 
excluding some upper limits that are shallower than other datapoints at similar $\nu$. 
The SEDs of the pulsars are normalized to $\Fradio$ measured at $\sim$ 1.5 GHz. 
The gray shaded region represents the power-law interpolation 
between the radio and X-ray fluences of FRB~200428A (i.e., spectral index $\beta = -0.60$--$-0.55$). 
The ranges of predicted optical emission by the model of IC in a pulsar magnetosphere (Y19) 
and by the model of a magnetar blastwave (B20) are shown with downward arrows 
at log$_{10} \nu$ [Hz] $\sim$ 15 (left and right, respectively). 
\label{fig:sed} 
}
\end{figure}

The fluences of the 11 radio bursts detected by FAST 
and discussed in this study are smaller than the typical fluence 
of FRBs detected by other radio telescopes, which is $\gtrsim 1$ Jy ms, 
and the small radio fluence makes the limit on $\Fopt/\Fradio$ less constraining. 
However, if a similar limit on optical fluence is obtained for a brighter radio burst, 
the limit on $\Fopt/\Fradio$ may put more strict constraints 
on the spectral energy distribution (SED) of the burst. 
For example, in case that the optical fluence limit of $< 0.029$ Jy ms is obtained 
for a radio burst with radio fluence $\Fradio = 1$, 10, and 100 Jy ms, 
the limit on $\Fopt/\Fradio$ is $< 10^{-1.5}, 10^{-2.5}$, and $10^{-3.5}$, respectively. 
Radio bursts with $\Fradio > 100$ Jy ms are indeed detected 
from several FRBs including repeating FRB~171019 \citep{Shannon:2018a, Kumar:2019a}. 
We compare the $\Fopt/\Fradio$ limits to theoretical 
and empirical models of an SED of a radio variable object below (figure~\ref{fig:sed}). 

It is difficult to robustly predict optical luminosity of a FRB 
as neither the origin nor the emission mechanism is known. 
However, it is possible that IC in a pulsar magnetosphere produce a short optical emission 
as bright as $\Fopt/\Fradio \sim 10^{-2}$ associated with an FRB as discussed in Y19. 
B20 also discussed that an optical emission 
of $\lesssim 10^{44}$ erg with a timescale $\lesssim 1$ sec can be produced 
when a blastwave from a magnetar impacts a hot wind bubble in the tail of a previous flare. 
Comparing the energy limit with the energy release of the brightest radio burst 
discussed in this study \citep[$4\times10^{38}$ erg,][]{Niu:2021a}, 
the prediction of the blastwave model corresponds to $\Fopt/\Fradio \sim 10^{-0.3}$. 
The current observational limits on $\Fopt/\Fradio$ obtained 
by the simultaneous observations are comparable 
to the brightest end of the possible range of optical emission predicted by the models. 
The parameter space of the models can be constrained once 
a similar limit on optical fluence as obtained 
in this study is achieved for a brighter radio burst. 

We also consider detectability of optical emission from an FRB 
in cases where FRBs have similar SED 
to that of galactic pulsars that are detected in optical passbands. 
Among pulsars that are detected in optical, the Crab pulsar is known 
to have bright optical emission with $\Fopt/\Fradio \geq 0.1$ 
when $\Fradio$ is measured at $\nu \sim 1.5$~GHz \citep[e.g.,][]{Buhler:2014a}. 
If an FRB typically has an SED which is similar to the Crab pulsar, 
our observations and some of the previous optical observations 
could have detected an FRB optical emission. 
Although other pulsars are fainter in optical, 
the Geminga has $\Fopt/\Fradio \gtrsim 0.01$ 
which can be detected with our observation for a radio burst with $\Fradio \gtrsim 5$ Jy ms. 
On the other hand, the Vela pulsar has $\Fopt/\Fradio \sim 10^{-6}$, 
with which optical emission from an FRB would be difficult to detect. 

The interpolation between the observed radio and X-ray fluences of FRB~200428A 
(a radio burst from a galactic magnetar SGR~1935+2154) 
is also considered as an SED template, 
although the spectral slope derived from the X-ray data alone 
does not agree with the interpolation \citep{Ridnaia:2021a}. 
The radio burst has fluence of 700 kJy ms and 1.5 MJy ms 
at $\sim$ 600 MHz and 1.4 GHz, respectively 
\citep{The-CHIME/FRB-Collaboration:2020a, Bochenek:2020a}. 
The X-ray flare that occurred simultaneously with FRB~200428A 
has fluence of $\sim$ 1 Jy ms at 100 keV \citep{Ridnaia:2021a}. 
By interpolating between the radio fluence at $\sim$ 600 MHz (1.4 GHz) 
and the X-ray fluence with a simple power-law, 
$F_\nu \propto \nu^\beta$, we obtain $\beta = -0.55$ ($-0.60$) 
which can be detected in optical 
with our observation when $\Fradio \gtrsim 100$ Jy ms. 

\section{Conclusions} 
\label{sec:conclusion}

We have conducted 24.4~fps optical observations of FRB~20190520B 
simultaneously with the monitoring observation by FAST. 
11 radio bursts are detected by FAST during the simultaneous observations, 
however no corresponding optical emission is found. 
The limits on optical emission within the image frames that correspond 
to the arrival times of the radio bursts are $<$ 0.068--0.149 Jy ms 
in terms of the fluence in the integration time of 40.9 ms, 
corrected for the foreground dust extinction in the MW. 
We also obtain the optical fluence limit of $< 0.029$ Jy ms 
by stacking the image frames that correspond to the radio bursts. 

Our fluence limit is deeper than those obtained by 
the previous simultaneous observations with sub-second time resolution 
for an optical emission with a duration $\gtrsim 0.1$ ms. 
Although the current limits on the optical to radio fluence ratio 
do not strictly constrain SED models of an FRB, 
some template SEDs based on optically detected pulsars 
and also a part of parameter spaces of the theoretical models of FRB optical emission 
by IC in a pulsar magnetosphere (Y19) and magnetar blastwave (B20) 
can be ruled out if a similar fluence limit as in our observation 
is obtained for a radio burst with $\Fradio \gtrsim 5$ Jy ms. 
With the progress in high-speed optical facilities and the discovery of various FRB sources, 
it is possible that optical emission from an FRB is found in the near future. 

\begin{acknowledgments} 
We would like to thank the anonymous referee for encouraging comments. 
This research was supported by JSPS KAKENHI 
Grant Number 20H01942, 18H05223, 18H01261 
and NSFC grant No. 11988101, 11725313. 
\end{acknowledgments} 


\software{astropy \citep{Astropy-Collaboration:2013a}, Source Extractor \citep{Bertin:1996a}, 
                scipy \citep{Virtanen:2020a} 
          }

\bibliography{reference_list}{}
\bibliographystyle{aasjournal}



\end{document}